\documentclass[11pt,a4paper]{article}
\pdfoutput=1
\usepackage{epsfig}
\usepackage{graphicx}
\usepackage[english]{babel}
\usepackage[utf8]{inputenc}
\usepackage{mathtools}
\usepackage{amsmath}    
\usepackage{jheppub}
\usepackage[english]{babel}
\usepackage[utf8]{inputenc}
\usepackage{amsmath}
\usepackage{graphicx}
\usepackage{mathtools}
\usepackage{braket}

\newcommand{\be}{\begin{equation}}
\newcommand{\ee}{\end{equation}}
\newcommand{\ben}{\begin{equation*}}
\newcommand{\een}{\end{equation*}}
\newcommand{\ba}{\begin{aligned}}
\newcommand{\ea}{\end{aligned}}

\title{Precursors and BRST Symmetry}

\author{Jan de Boer, Ben Freivogel, Laurens Kabir and Sagar F. Lokhande}
\affiliation{Institute for Theoretical Physics Amsterdam and Delta Institute for Theoretical Physics\\University of Amsterdam\\Science Park 904\\1098 XH Amsterdam, The Netherlands}

\emailAdd{J.deBoer@uva.nl}
\emailAdd{benfreivogel@gmail.com}
\emailAdd{laurenskabir@gmail.com}
\emailAdd{sagar.f.lokhande@gmail.com}

\abstract{In the AdS/CFT correspondence, bulk information appears to be encoded in the CFT in a redundant way. A local bulk field corresponds to many different non-local CFT operators (precursors). We recast this ambiguity in the language of BRST symmetry, and propose that in the large $N$ limit, the difference between two precursors is a BRST exact and ghost-free term. Using the BRST formalism and working in a simple model with global symmetries, we re-derive a precursor ambiguity appearing in earlier work. Finally, we show within this model that this BRST ambiguity has the right number of parameters to explain the freedom to localize precursors within the boundary of an entanglement wedge order by order in the large $N$ expansion.}

\begin{document}
\maketitle

\section{Introduction}
The AdS/CFT correspondence is the most precise non-perturbative definition of quantum gravity. A central problem is how local bulk physics emerges from CFT data. This question has been studied extensively and is reasonably well-understood at large $N$, for small perturbations around vacuum AdS  \cite{Balasubramanian:1998sn, Banks:1998dd}. In this limit, a bulk field $\Phi$ at a point $X$ is defined by integrating a local CFT operator $\mathcal{O}$ over the boundary with an appropriate smearing function $K$ \cite{Hamilton:2006az}:
\be
\Phi(X)= \int dt \, d^{d-1}x  \; K(X|t,{x}) \mathcal{O}({x}) + O\bigg(\frac{1}{N}\bigg).
\ee
This CFT operator can subsequently be time evolved to a single timeslice using the CFT Hamiltonian, which gives a non-local operator $P$ in the CFT corresponding with the field $\Phi(X)$ in the bulk.  This type of operator is called a `precursor' \cite{Polchinski:1999yd, Giddings:2001pt, Freivogel:2002ex}.\\

The study of precursors is fundamental to understanding a concrete realization of holography. There are several unresolved questions one can ask, such as how to construct precursors that correspond to bulk fields behind a black hole horizon.
Here we focus on two particular puzzles that are related to each other.
At large N, bulk locality requires the precursor to commute with all local CFT operators at a fixed time, while basic properties of quantum field theory demand that only trivial operators can commute with all local operators at a given time \cite{Almheiri:2014lwa}. Another is that a local bulk operator corresponds with many different precursors with different spatial support in the CFT, because the bulk field can be reconstructed in a particular spatial region of the CFT as long as it is contained in the corresponding entanglement wedge of that region. 

Both of these apparent paradoxes can be resolved by requiring that different precursors are not equivalent as true CFT operators \cite{Almheiri:2014lwa}. In particular, the difference between two precursors corresponding to the same bulk field seems to have no clear physical meaning, and must act trivially some class of states. In what follows, we will refer to this perplexing feature as the `precursor ambiguity'. \\

In \cite{Almheiri:2014lwa} and \cite{Mintun:2015qda} some progress was made in giving a guiding principle for constructing the ambiguity between two precursors corresponding to the same bulk field. The former approach recasts the AdS/CFT dictionary in the language of quantum error correction (QEC). From this viewpoint, the ambiguity is an operator which acts trivially in the code subspace of QEC, which in this case is naturally thought of as the space of states dual to low-energy excitations of the bulk. The latter work, on the other hand, proposed that gauge symmetry in the CFT can give a prescription to construct the precursor ambiguity. Moreover, they claimed that the code subspace is the full space of gauge invariant states.\\

In this paper, we start in section \ref{sec:conjecture} by proposing the language of BRST symmetry as a tool for making the precursor ambiguity concrete. In section \ref{sec:BRST}, we show that this approach nicely reduces to an already identified precursor ambiguity in the presence of a global $SO(N)$ symmetry \cite{Mintun:2015qda}. Furthermore, it has the added benefit that it generalizes to arbitrary gauge theories at any $N$. In section \ref{sec:localizebulk} we show in a particular toy model how this precursor ambiguity has the right number of parameters  to enable us to localize precursors in the boundary of the entanglement wedge order by order in 1/N. 

\section{Proposal: Precursor Ambiguities from BRST}\label{sec:conjecture}
In most of the known examples of holography, the boundary theory has some gauge symmetry. The presence of these `unphysical' degrees of freedom renders the naive path integral for gauge theories divergent. One approach to deal with these problems while covariantly quantizing the gauge theory is the BRST formalism \cite{Becchi:1975nq, Tyutin:1975qk}. The rough idea is to replace the original gauge symmetry with a global symmetry, by enlarging the theory and introducing additional fields. This new rigid symmetry, the BRST symmetry, will still be present after fixing the gauge. Since the generator of the BRST symmetry $Q_{\text{BRST}}$ is nilpotent of order two, we can construct its cohomology which will describe the gauge invariant observables of the original theory. \\

We propose that the natural framework to understand precursor ambiguities is the language of BRST symmetry. In particular, we claim that if $P_1$ and $P_2$ are two precursors in the large $N$-limit corresponding with the same local bulk field $\Phi(X)$, then $P_1 - P_2 = \mathcal{O}$ where
\begin{itemize}
\item $\mathcal{O}$ is BRST exact: $\mathcal{O}=\{Q_{\text{BRST}},\tilde{\mathcal{O}}\}$ 
\item $\mathcal{O}$ does not contain any (anti-)ghosts.
\end{itemize}
By construction this leaves any correlation function of gauge invariant operators in arbitrary physical states invariant
\be
\langle \mathcal{O}_1 \cdots \mathcal{O}_i \cdots \mathcal{O}_n\rangle= \langle \mathcal{O}_1 \cdots (\mathcal{O}_i+\{Q_{\text{BRST}},\tilde{\mathcal{O}}\} ) \cdots \mathcal{O}_n\rangle
\ee
since $[Q_{\text{BRST}},\mathcal{O}_i]=0$ for a gauge invariant operator $\mathcal{O}_i$, and $Q_{\text{BRST}}|\psi\rangle=0$ for a gauge invariant state $|\psi\rangle$. \\

As an example, we will show in section \ref{sec:BRST} that in the case of $N$ free scalars with a global ${\rm SO}(N)$ symmetry, we can reproduce the results of \cite{Mintun:2015qda}. That means, there exists an operator $\tilde{\mathcal{O}}$ such that
\be
\{Q_{\text{BRST}},\tilde{\mathcal{O}}\}\sim L^{ij} A^{ij}
\ee
where $L^{ij}$ is the generator of the ${\rm SO}(N)$ symmetry, and $A^{ij}$ is any operator in the adjoint.\\

Note that while the BRST ambiguity is well-defined for any gauge theory and even at finite $N$, the notion of bulk locality only makes sense perturbatively in $1/N$. In order to connect the abstract BRST ambiguity to concrete equivalences between different CFT operators, we need to make use of the large $N$ expansion. Thus the precursor ambiguity we find is valid within states where the number of excitations is small compared to $N$.

\section{BRST Symmetry of $N$ Real Scalars}\label{sec:BRST}
%
In this section we will apply the BRST formalism to a theory of $N$ real scalars. The Lagrangian for this gauge theory in the covariant gauge is given by
\be
\mathcal{L}=-\frac14 (F^a_{\mu \nu})^2 + \frac12 D^\mu \phi^i D_\mu \phi_i + \frac{\xi}{2} (B^a)^2 +B^a \partial^\mu A_\mu^a + \partial^\mu\bar{c}^a  (D_\mu^{}c)^a
\ee
where the auxiliary field $B^a$ can be integrated out using $\xi B^a=-\partial^\mu A_\mu^a$.
We take the $\phi^i$ in the fundamental representation of ${\rm SO}(N)$, while the ghost $c^a$, anti-ghost $\bar{c}^a$ and the gauge field $A_\mu^a$ are in the adjoint. The (anti-)ghosts are scalar fermion fields.
The covariant derivatives are given by
\be
(D_\mu^{} c)^a = \partial_\mu c^a + g f^{abc} A^b_\mu c^c
\ee
and
\be
(D_\mu^{} \phi)^i = \partial_\mu \phi^i - i g A_\mu^a (T^a)_{ij}\phi^j.
\ee
Note that $D_\mu^{} \phi^i$ is real since the matrices $(T^a)_{ij}$ are purely imaginary for ${\rm SO}(N)$.
The field strength $F$ is given by
\be
F^a_{\mu \nu}=\partial_\mu A_\nu^a - \partial_\nu A^a_\mu + g f^{abc}A^b_\mu A^c_\nu.
\ee
This Lagrangian is invariant under the following BRST symmetry:
\begin{align}
\delta_B A^a_\mu&=\epsilon (D_\mu c)^a \nonumber\\
\delta_B \phi^i &= i g \epsilon c^a (T^a)_{ij}\phi^j \nonumber\\
\delta_B c^a &= -\frac12  g \epsilon f^{abc} c^b c^c \\
\delta_B \bar{c}^a &= \epsilon B^a\nonumber\\
\delta_B B^a &=0. \nonumber
\label{eq:BRSTtransfos}
\end{align}
where $\epsilon$ is a constant Grassmann parameter.

\subsection{The BRST Charge}
In order to compute the BRST charge, we start by constructing the Noether current associated to this symmetry
\be
J^\mu = \sum_\alpha \frac{\delta \mathcal{L}}{\delta( \partial_\mu \Phi_\alpha)} \delta_B \Phi_\alpha
\ee
where the sum runs over all possible fields in the Lagrangian.
The BRST charge is then defined via
\be
Q_B=\int d^{d-1}x \; J^0_B
\ee
and generates the BRST transformations on the fields via 
\be
\delta_B \Phi_\alpha=\epsilon [\Phi_\alpha,Q_{\text{BRST}}]_\pm.
\ee
Let's start by computing the variations and defining the conjugate momenta
\begin{align}
\frac{\delta \mathcal{L}}{\delta( \partial_\mu \phi^i)}&= D_\mu \phi^i\qquad & \Pi^i &\equiv D_0 \phi^i \qquad &[\phi^i(x), \Pi^j(y)]&= \delta^{ij}\delta^{(d-1)}({x} - {y})\\
\frac{\delta \mathcal{L}}{\delta( \partial_\mu c^a)}&= (\partial_\mu \bar{c})^a \qquad & \pi_c^a &\equiv (\partial_0 \bar{c})^a \qquad &\{c^a(x),\pi_c^b(y)\}&=\delta^{ab}\delta^{(d-1)}({x} - {y})\\
\frac{\delta \mathcal{L}}{\delta( \partial_\mu \bar{c}^a)}&= (D_\mu c)^a \qquad & \pi_{\bar{c}}^a &\equiv (D_0 c)^a \qquad &\{\bar{c}^a(x),\pi_{\bar{c}}^b(y)\}&=\delta^{ab}\delta^{(d-1)}({x} - {y})
\end{align}
and finally for the gauge field
\be
\frac{\delta \mathcal{L}}{\delta( \partial_\mu A^a_\nu)}=- F^a_{\mu\nu} + \eta_{\mu \nu}B^a \qquad \Pi^a_\nu\equiv -F^a_{0\nu} + \eta_{0 \nu}B^a 
\ee
with commutation relation
\be
[A^a_\mu(x),\Pi^b_\nu(y)]= \eta_{\mu \nu}\delta^{ab} \, \delta^{(d-1)}({x} - {y}).
\ee
That gives the following Noether current
\be
J^\mu= \left( - F^a_{\mu \nu} +\eta_{\mu \nu}B^a\right) (D_\nu c)^a+ i g D_\mu \phi^i c^a (T^a)_{ij}\phi^j - \frac12 g  (\partial_\mu \bar{c}^a)f^{abc} c^b c^c+ (D_\mu c)^a B^a.
\ee
The BRST charge is then given by
\begin{align}
Q_{\text{BRST}}&=\int dx^{d-1} \;  \Pi^a_\nu (D_\nu c)^a + i g \Pi^i c^a (T^a)_{ij}\phi^j - \frac12 g  f^{abc} \pi^a_c  c^b c^c + B^a\pi_{\bar{c}}^a \\
&=\int dx^{d-1} \;  \Pi^a_\nu (\partial_\nu c)^a -  g f^{abc} A_\nu^b\Pi^c_\nu   c^a+ig  \Pi^i c^a (T^a)_{ij}\phi^j - \frac12 g  f^{abc} \pi^a_c  c^b c^c + B^a\pi_{\bar{c}}^a. \nonumber
\end{align}
We can define the generators of the ${\rm SO}(N)$ symmetry, as the Noether currents associated with the gauge transformations. The current has two contributions, one from the Yang-Mills parts $F^2$ and one from the matter part $(D \phi)^2$:
\be
J^a_{\text{matter}}\equiv i\, \Pi^i  (T^a)_{ij}\phi^j \qquad J^a_{\text{gauge}}\equiv - f^{abc}A^b_\mu \Pi^c_{\mu}  
\ee
\be
J^a \equiv \left( J^a_{\text{matter}} + J^a_{\text{gauge}}\right) .
\ee
This finally leads to the the BRST charge:
\be
Q_{\text{BRST}}=\int dx^{d-1} \; \left( g c^a J^a - \frac12 g  f^{abc} \pi^a_c  c^b c^c+B^a \pi_{\bar{c}}^a +\Pi^a_\nu (\partial_\nu c)^a \right).
\ee

\subsection{Reduction to a Global ${\rm SO}(N)$ Symmetry}
In order to connect with previous work on precursors \cite{Mintun:2015qda}, we are interested in degrading the $SO(N)$ gauge symmetry to a global symmetry. One crude way of accomplishing this, is by setting the gauge fields $A^a_\mu=0$ (and also $B^a=0$ since $B^a \sim \partial^\mu A^a_\mu$). In this case, the ghosts become quantum mechanical (position independent) and the BRST charge reduces to
\be
Q_{\text{BRST}}=\int dx^{d-1}  g c^a J^a - \frac12 g  f^{abc} \pi^a_c  c^b c^c  \qquad J^a=i\, \Pi^i  (T^a)_{ij}\phi^j
\ee
where the global $SO(N)$ generator is given by $L^a=\int d^{d-1} x \, J^a(x)$. \\

Now consider an operator of the form $\pi_c^a \, \mathcal{O}^a$ and compute the anti-commutator with the BRST charge:
\begin{align}
\{ Q_{\text{BRST}},\pi^d_c \mathcal{O}^d\}&=\int dx^{d-1}  g \{  c^a J^a ,\pi^d_c \mathcal{O}^d\} -  \frac12 g  f^{abc} \{\pi^a_c  c^b c^c,\pi^d_c \mathcal{O}^d\}\\
&=g \int dx^{d-1}  \mathcal{O}^a J^a =g L^a \mathcal{O}^a
\label{eq:MPRAmbiguity}
\end{align}
where we used that the generator of global ${\rm SO}(N)$ transformations rotates the operator $\mathcal{O}$ as $[J^a , \mathcal{O}^b]=f^{abc} \mathcal{O}^c $.
This expression is BRST exact by construction, and ghost-free. Adding this to a CFT operator will have no effect whatsoever within correlation functions in physical states. It is exactly the precursor ambiguity found in \cite{Mintun:2015qda}.

\pagebreak

%

\section{Localizing Precursors in a Holographic Toy Model}\label{sec:localizebulk}
In the previous section, we computed the ambiguous part of the precursors as a BRST exact and ghost-free operator. This ambiguity can be viewed as the redundant, quantum error correcting part of the precursors. Once it has been identified, the physical information contained in the precursors becomes clear. In this section we will study the particular ambiguity \eqref{eq:MPRAmbiguity} in a toy model. We will show that this ambiguity has the structure of an HKLL series, and that it contains enough freedom to localize bulk information in a particular region of the CFT by setting the smearing function to zero in that region.
\subsection{The Model}
The model is a CFT containing $N$ free scalar fields in $1+1$ spacetime dimensions:
\begin{equation}
 \mathcal{L} = \sum\limits_{i=1}^N -\frac{1}{2} \, \partial_\mu \, \phi^i \, \partial^\mu \, \phi^i .
\end{equation}
It was first considered by \cite{Mintun:2015qda} and refined in \cite{Freivogel:2016zsb}. There is a $\Delta=2$ primary operator $\mathcal{O}=\partial_\mu \phi^i \, \partial^\mu \phi^i$ which we take to be dual to a massless scalar $\Phi$ in $AdS_ {2+1}$. \\ 

Following \cite{Mintun:2015qda} and \eqref{eq:MPRAmbiguity}, the precursor ambiguity is given by $L^{ij} A^{ij}$ where $A^{ij}$ is any operator in the adjoint of ${\rm SO}(N)$ and $L^{ij}$ is the generator of global ${\rm SO}(N)$ transformations. Note that we only kept the global part of the ${\rm SO}(N)$ transformations by setting $A_\mu^a=B^a=0$ in the full gauge theory discussed in section \ref{sec:BRST}.\\ 

Expanding the boundary field $\phi$ in terms of left/right-moving creation and annihilation modes, one can compute the generator of global rotations
\begin{equation}
L^{ij} = \int \frac{d k}{2 \, k} \, \bigg( \alpha^{\dagger}{}^{[i}_k \, \alpha_k^{j]} +  \tilde{\alpha}^{\dagger}{}^{[i}_k \, \tilde{\alpha}_k^{j]} \bigg)
\end{equation}
where the tilde denotes a right-moving polarization of the creation or annihilation modes and any zero modes are left out. 
If there is no confusion what momentum a given mode has, we will omit the subscript $k$.

\subsection{Precursor Ambiguity and Bulk Localization Perturbatively in $1/N$}
The bulk field $\Phi$ in global $AdS_3$ can be constructed at large $N$ by smearing quadratic operators of the form $\mathcal{O}\sim {\alpha_{k} \tilde{\alpha}_{k'} }$ over a particular region of the CFT \cite{Hamilton:2006az}:
\be
\Phi(X)=\int d^2 x \, K_1(X|x) \, \mathcal{O}(x) + O\bigg(\frac{1}{\sqrt{N}}\bigg)
\label{eq:HKLL}
\ee
where the smearing function $K$ obeys the bulk free wave equation
\be
\Box_{AdS_3} K_1(X|x)=0.
\ee
This procedure correctly reproduces the bulk two-point function. The precursor can be obtained from \eqref{eq:HKLL} by time evolving the CFT operator to a single timeslice. Extending the HKLL procedure perturbatively in $1/N$ will look schematically as follows \cite{Kabat:2011rz, Heemskerk:2012mn}:
\be
\Phi^{}(X)= \int K_1 \mathcal{O} +\frac{1}{\sqrt{N}} \iint K_2 \mathcal{O} \mathcal{O} + O\bigg(\frac{1}{N}\bigg)
\label{eq:interactingHKLL}
\ee
where the expansion parameter is $1/\sqrt{N}$ instead of $1/N$ because we are dealing with a vector-like theory \cite{Klebanov:2002ja}.\\

In \cite{Freivogel:2016zsb} it was shown that, at leading order in $1/N$, the spatial support of the smearing function $K_1$ (and hence the information of the bulk field) can be localized in a particular Rindler wedge of the CFT due to an ambiguity in the smearing function. This freedom can be understood by noting that the term $ \alpha_{k_1}^{\dagger i} \, \tilde{\alpha}_{k_2}^i $ can be added to $\mathcal{O}$ within two-point functions since it annihilates the vacuum in both directions. While this two-parameter family of freedom is enough to localize the bulk field at leading order in $N$, one can see that it generically will be insufficient to set $K_2$ to zero in particular region, because this requires a four-parameter family of freedom. Since changing the smearing function corresponds with picking a different precursor, we would like to identify the aforementioned freedom in the smearing function with the precursor ambiguity. In what follows, we will explain how the precursor ambiguity $L^{ij} A^{ij}$ has enough freedom to localize bulk information order by order in $1/N$.\\

Start by considering the following quadratic (adjoint) operator
\begin{equation}
A^{ij}_2 \equiv \alpha_{k_1}^{\dagger i} \, \tilde{\alpha}_{k_2}^j .
\end{equation}
A possible ambiguity of the precursor will be given by $L^{ij} \, A^{ij}_2$. 
Normal ordering yields
\begin{align}
\begin{split}
\frac{1}{N^\frac32}L^{ij} \, A^{ij}_2 &= \frac{1}{N^\frac32} \int \frac{d k}{2 \, k} \,  \bigg( \alpha^{\dagger}{}^{[i}_k \, \alpha_k^{j]} +  \tilde{\alpha}^{\dagger}{}^{[i}_k \, \tilde{\alpha}_k^{j]}  \bigg) \, \alpha_{k_1}^{\dagger i} \, \tilde{\alpha}_{k_2}^j \\
&=  \frac{(1-N)}{N^\frac32} \, \alpha^{\dagger i}_{k_1} \, \tilde{\alpha}^i_{k_2} +\frac{1}{\sqrt{N}} \frac{\alpha^{\dagger}{}^{i}_{k_1} \, L^{ij} \, \tilde{\alpha}^j_{k_2}}{N} \\
&\sim \mathcal{O} + \frac{1}{\sqrt{N}} \mathcal{O}\mathcal{O}
\end{split}
\label{eq:quadraticambiguity}
\end{align}
where $\mathcal{O}$ denotes an operator quadratic in the $\alpha$'s and normalized by $1/\sqrt{N}$ such that it is $O(1)$ in $N$-scaling. Note that the LHS of \eqref{eq:quadraticambiguity}, by construction, is zero in physical states (and hence can be added to the precursor without changing any of its correlation functions). \\

The piece quadratic in the $\alpha's$ in \eqref{eq:quadraticambiguity} is exactly the ambiguity needed to localize the precursor in the CFT to leading order in $N$, as was shown in detail in \cite{Freivogel:2016zsb}. One can now also see that one generically needs a four-parameter ambiguity if we want to be able to set $K_2$ in \eqref{eq:interactingHKLL} to zero in certain regions. Even though the term $\mathcal{O}\mathcal{O}/\sqrt{N}$ in \eqref{eq:quadraticambiguity} has the right structure to fit in the HKLL series, it does not have enough freedom to set $K_2$ to zero (it has only 2 free parameters, while we need 4). It can be done, however, by constructing a new operator which annihilates $SO(N)$-invariant states and is quartic in the $\alpha$'s:
\begin{equation}
 A^{ij}_4 \equiv A_2^{ij} - \frac{1}{N} \,  A_2^{ij} \, \alpha^{\dagger \, m}_{k_3} \, \alpha^m_{k_4} .
\end{equation}
The ambiguity in the precursor to order $\frac{1}{\sqrt{N}}$ is then given by $L^{ij} \, A^{ij}_4$. Normal ordering yields
\begin{equation}
L^{ij} \, A^{ij}_4 = L^{ij} \, A^{ij}_2 + T_4 + T_6
\end{equation}
where
\begin{equation}
T_4 =  \alpha^{\dagger \, i}_{k_1} \, \alpha^{\dagger \, i}_{k_3} \, \tilde{\alpha}^m_{k_2} \, \alpha^m_{k_4} - \alpha^{\dagger \, i}_{k_3} \, \tilde{\alpha}^i_{k_2} \, \alpha^{\dagger \, m}_{k_1} \, \alpha^m_{k_4} + (1-N) \, \alpha^{\dagger \, i}_{k_1} \, \tilde{\alpha}^i_{k_2} \, \alpha^{\dagger \, m}_{k_3} \, \alpha^m_{k_4}   
\end{equation}
\begin{align}
 \begin{split}
T_6 &= \alpha^{\dagger \, i}_k \, \alpha^{\dagger \, i}_{k_1} \, \alpha^{\dagger \, m}_{k_3} \, \tilde{\alpha}^j_{k_2} \, \alpha^j_k \, \alpha^m_{k_4} - \alpha^{\dagger \, j}_k \, \alpha^{\dagger \, i}_{k_1} \, \alpha^{\dagger \, m}_{k_3} \, \tilde{\alpha}^j_{k_2} \, \alpha^i_k \, \alpha^m_{k_4} \\
&\hspace{3mm}+ \tilde{\alpha}^{\dagger \, i}_k \, \alpha^{\dagger \, i}_{k_1} \, \alpha^{\dagger \, m}_{k_3} \, \tilde{\alpha}^j_{k_2} \, \tilde{\alpha}^j_k \, \alpha^m_{k_4} - \tilde{\alpha}^{\dagger \, j}_k \, \alpha^{\dagger \, i}_{k_1} \, \alpha^{\dagger \, m}_{k_3} \, \tilde{\alpha}^j_{k_2} \, \tilde{\alpha}^i_k \, \alpha^m_{k_4}  
 \end{split}
\end{align}
and repeated momenta are integrated over appropriately. By $T_4$ we denote the ambiguity to quartic order in $L^{ij} \, A^{ij}_4$ and similarly with $T_6$ to hexic order. 
As before, $T_4$ and $T_6$ scale the same with respect to $N$ in any gauge invariant state. Also they do not contribute in three-point functions of the bulk field. \\

Again we find that all the terms nicely arrange themselves in the right structure of an HKLL series
\be
L^{ij} A_4^{ij}\sim \mathcal{O} + \frac{1}{\sqrt{N}} \mathcal{O}\mathcal{O} + \frac{1}{N} \mathcal{O}\mathcal{O}\mathcal{O} 
\label{eq:quarticambiguity}
\ee
where $\mathcal{O}$ schematically denotes an operator quadratic in the $\alpha$'s and normalized by $1/\sqrt{N}$ such that it is $O(1)$ in $N$-scaling. The main difference with $L^{ij} A_2^{ij}$ is that the term quartic in the $\alpha$'s now gets a contribution from $T_4$, which does have four independent parameters, and hence has enough freedom to localize the smearing function $K_2$.\\



Doing so also introduced a term like $\alpha^6$. The connected piece of this will be down in $1/N$ relative to $\alpha^4$. If $T_4$ fixes the ambiguity at order $1/\sqrt{N}$, $T_6$ will contribute towards fixing it at order $1/N$. Thus, by choosing a proper operator $A^{ij}$, we will be able to fix the ambiguity in the precursor to any order in $1/N$ perturbatively.\\
 
We can now summarize how this recursive procedure works to localize bulk information order by order in $N$. When the operator we want to smear $A^{ij}_2$ is quadratic, the ambiguity in the precursor to the quadratic order is given by $(1-N) \, \alpha^{\dagger i}_{k_1} \tilde{\alpha}^j_{k_2}$. These modes are labeled by two different momenta. Since we are working in two spacetime dimensions, they are able to fix all the ambiguity in the precursor up to quadratic level. \\

But fixing the quadratic level, introduces a quartic piece: $\alpha^{\dagger}{}^{i}_{k_1} \, L^{ij} \, \tilde{\alpha}^j_{k_2}$.  This piece has insufficient freedom to localize the precursor up to $1/\sqrt{N}$ effects. To fix the ambiguity to the quartic level, one introduces a quartic ambiguity $L^{ij} A^{ij}_4$. This gives a piece $T_4$ which has four independent momenta and hence can now fix any ambiguity in the precursor up to quartic order. However, doing so also introduced a hexic piece $T_6$. This hexic term makes the precursor ambiguous to order six. We can repeat the procedure, smear a different $A^{ij}$ and then fix the ambiguity in the precursor up to order six.\\

Surprisingly, each term at a higher order is $\frac{1}{\sqrt{N}}$ relative to the current order. Hence, this procedure can be carried out order by order in $\frac{1}{\sqrt{N}}$ and thus fixes all the ambiguity in the interacting HKLL series in this toy model. While it is not explicitly demonstrated in this paper, a similar story should hold when the matter fields are in the adjoint.\\

One should note that, while the quadratic and quartic piece in the ambiguity \eqref{eq:quadraticambiguity} (and similarly for the quartic and hexic piece in the ambiguity \eqref{eq:quarticambiguity}) have the correct `naive' $N$-scaling ($\alpha \sim N^\frac14$) to be arranged in an HKLL series, their real $N$-scaling is the same. This means that neither term in \eqref{eq:quadraticambiguity} or \eqref{eq:quarticambiguity} is smaller compared to the other.  For clarity, we will elaborate on this a bit more in the next section \ref{sec:Nscaling}.

\subsection{$N$-Scaling}\label{sec:Nscaling}
Within physical states, both terms on the RHS of \eqref{eq:quadraticambiguity} will be equal and opposite. In particular, they must have the same $N$-scaling (in contrary to what was claimed in \cite{Mintun:2015qda}), even though naive $N$-counting would suggest otherwise. In order to explicitly see that both terms have the same $N$-scaling in $SO(N)$-invariant states, we pick the following three states and label the operators as follows:
\begin{center}
 \begin{tabular}[h]{|c|c|}
 \hline
 States & Operators\\
\hline \hline
$| \psi_1' \rangle = \frac{1}{\sqrt{N}} \, \alpha_{k_3}^{\dagger m} \, \alpha_{k_4}^{\dagger m} \, | 0 \rangle $ & $\mathcal{O}_1=\alpha_{k_1}^{\dagger i} \, L^{ij} \, \tilde{\alpha}_{k_2}^{j}/ N^\frac32$ \\
\hline
$| \psi_1'' \rangle = \frac{1}{\sqrt{N}} \, \tilde{\alpha}_{k_3}^{\dagger m} \, \tilde{\alpha}_{k_4}^{\dagger m} \, | 0 \rangle$ & $\mathcal{O}_2= \, \alpha_{k_1}^{\dagger i} \, \tilde{\alpha}_{k_2}^i/\sqrt{N}$ \\
\hline
$| \psi_2 \rangle = \frac{1}{\sqrt{N}} \, \tilde{\alpha}_{k_5}^{\dagger m} \, \alpha_{k_6}^{\dagger m} \, | 0 \rangle$ &\\
\hline
\end{tabular}
\end{center}
In order to assign a $N$-scaling to $\mathcal{O}_2$, one could check its two-point function. However, since this operator has vanishing two-point functions, we investigate the three-point function and find that it goes like $1/\sqrt{N}$. This justifies us to call assign an $O(1)$ $N$-scaling to $\mathcal{O}_2$. We will estimate the size of $\mathcal{O}_1$ and $\mathcal{O}_2$ in the subspace spanned by the three states above. Let us denote the matrix elements of an arbitrary operator $\mathcal{O}$ in the above subspace as
\begin{center}
$\mathcal{O} = \begin{pmatrix}
 \langle \psi_1' | \mathcal{O} | \psi_1' \rangle & \langle \psi_1' | \mathcal{O} | \psi_1'' \rangle & \langle \psi_1' | \mathcal{O} | \psi_2 \rangle\\
 \langle \psi_1'' | \mathcal{O} | \psi_1' \rangle & \langle \psi_1'' | \mathcal{O} | \psi_1'' \rangle & \langle \psi_1'' | \mathcal{O} | \psi_2 \rangle \\
 \langle \psi_2 | \mathcal{O} | \psi_1' \rangle & \langle \psi_2 | \mathcal{O} | \psi_1'' \rangle & \langle \psi_2 | \mathcal{O} | \psi_2 \rangle
\end{pmatrix}.$
\end{center}
Then we get the following matrix elements for $\mathcal{O}_1$ and $\mathcal{O}_2$
\begin{align}
 \mathcal{O}_1 =\frac{1}{\sqrt{N}} \begin{pmatrix}
0 & 0 & 1 \\
0 & 0 & 0 \\
0 & 1 & 0        
\end{pmatrix} \qquad
\mathcal{O}_2 =\frac{1}{\sqrt{N}}  \begin{pmatrix}
0 & 0 & 1 \\
0 & 0 & 0 \\
0 & 1 & 0        
\end{pmatrix}.
\end{align}

We can see that both the pieces in $L^{ij} \, A^{ij}_2$ scale in the same way with respect to $N$, as expected. Naively, one could expect the part quartic in the $\alpha$'s to be down to part quadratic in the $\alpha$'s by a factor $1/\sqrt{N}$. For these particular operators that doesn't happen, because the disconnected piece in $\mathcal{O}_1$ enhances its $N$-scaling. \\

Applying similar arguments to \eqref{eq:quarticambiguity}, we conclude $T_6$ must have the same $N$-scaling as $T_4$. Again, the reason why this does not agree with naive $N$-scaling, is due to the contribution from the disconnected piece in $T_6$. 

\section{Outlook}

In this paper we have presented preliminary evidence that precursors are related to BRST invariance and hence to 
the underlying gauge symmetry of the field theory. There are several interesting follow-up directions to explore.
One could for example study precursors in the toy model in non-trivial states (such as thermal states), but
more importantly, one would like to generalize the construction to a proper gauge theory with local gauge invariance.
Perhaps the simplest example of a field theoretic precursor ambiguity is to consider the field theoretic dual of the bulk
operator one obtains by integrating a bulk field over a symmetric minimal surface. Such operators were studied in \cite{Czech:2016xec,
deBoer:2016pqk}, and to lowest order in the $1/N$ expansion in the field theory for a bulk scalar they are given by
\begin{equation}
 Q_{\cal O}(x,y) 
=C \int_{D(x,y)} \!\!\!\!\!d^d \xi\, \left( \frac{(y-\xi)^2(\xi-x)^2 }{-(y-x)^2} \right)^{\frac{(\Delta_{\cal O}-d)}{2}}\ 
\langle {\cal O}(\xi)\rangle\,
\end{equation}
where the integral is over the causal diamond $D(x,y)$ with past and future endpoints $x$ and $y$, and $\Delta_{\cal O}$
is the scaling dimension of the primary operator ${\cal O}$. The constant $C$ is a normalization constant which at
this point is arbitrary. The past light-cone of $y$ and the future light-cone of $x$ intersect at a sphere $B$, which is
the boundary of the bulk minimal surface.\\

If the field theory is defined on $S^{d-1}\times \mathbb R$, then there are two equivalent choices of causal diamonds 
for a given symmetric minimal surface. Together, they contain a full Cauchy slice for the field theory.
Hence, there are two inequivalent boundary representations of the same bulk operator, 
and the difference between these two is an example of a precursor ambiguity. We would therefore like to conjecture
that there exists an operator $Y$ such that
\begin{align} \label{aux1}
\{Q_{\rm BRST}, Y\} & = \int_{D(x,y)} \!\!\!\!\!d^d \xi\, \left( \frac{(y-\xi)^2(\xi-x)^2 }{-(y-x)^2} \right)^{\frac{(\Delta_{\cal O}-d)}{2}}\ 
\langle {\cal O}(\xi)\rangle\ \\
&- \int_{\bar{D}(\bar{x},\bar{y})} \!\!\!\!\!d^d \bar{\xi}
\, \left( \frac{(\bar{y}-\bar{\xi})^2(\bar{\xi}-\bar{x})^2 }{-(\bar{y}-\bar{x})^2} \right)^{\frac{(\Delta_{\cal O}-d)}{2}}\ 
\langle {\cal O}(\bar{\xi})\rangle\ + { O}(1/N)
\end{align}
Here, the second complimentary causal diamond is denoted by $D(\bar{x},\bar{y})$ with past and future endpoints $\bar{x}$ and $\bar{y}$.
It would be very interesting to construct an operator $Y$ for which (\ref{aux1}) holds, and we hope to come back to this
in the near future.


\acknowledgments
We thank Vladimir Rosenhaus for helpful discussions. This work is part of the Delta ITP consortium, a program of the Netherlands Organisation for Scientific Research (NWO) that is funded by the Dutch Ministry of Education, Culture and Science (OCW). SFL would like to acknowledge financial support from FOM, which is part of the NWO.

\bibliographystyle{ytphys}
\bibliography{ref}

\end{document}